\documentclass[3p,times]{elsarticle}
\usepackage[bookmarks=false]{hyperref}
    \hypersetup{colorlinks,
      linkcolor=blue,
      citecolor=blue,
      urlcolor=blue}

\usepackage{amssymb}

\usepackage[figuresright]{rotating}
\usepackage{graphicx}
\usepackage{dcolumn}
\usepackage{bm}
\usepackage{xcolor} 
\usepackage{amsmath}

\AtBeginDocument{
  \setlength{\abovedisplayskip}{6pt}
  \setlength{\belowdisplayskip}{6pt}
  \setlength{\abovedisplayshortskip}{6pt}
  \setlength{\belowdisplayshortskip}{6pt}
}

\begin{document}
\begin{frontmatter}




\title{\textit{LBFAST}: A Lightweight Moment-Represented Lattice Boltzmann Solver for Multi-GPU Architectures}


\author[a]{Marco Lauricella\corref{cor1}}
\ead{marco.lauricella@cnr.it}

\cortext[cor1]{Corresponding author.}

\author[b]{Andrea Montessori}
\author[c]{Giorgio Amati}
\author[d]{Filippo Spiga}
\author[a]{Adriano Tiribocchi} 
\author[a]{Massimo Bernaschi} 
\author[a,e,f]{Sauro Succi} 

\address[a]{Istituto per le Applicazioni del Calcolo, Consiglio Nazionale delle Ricerche, Via dei Taurini 19, Rome, 00185, Italy}
\address[b]{Department of Civil, Computer Science and Aeronautical Technologies Engineering, Roma Tre University, Via Vito Volterra, Rome, 00146, Italy}
\address[c]{HPC Department, CINECA, Rome 00185, Italy}
\address[d]{NVIDIA Development UK Ltd, Milton Hall, Ely Rd, Milton, Cambridge CB24 6WZ, United Kingdom}
\address[e]{Center for Life Nano- \& Neuro-Science, Fondazione Istituto Italiano di Tecnologia, Viale Regina Elena 295, Rome, 00161, Italy}
\address[f]{Department of Physics, Harvard University, 17 Oxford St., Cambridge, 02138, MA, USA}

\begin{abstract}
We present \textit{LBFAST}, a GPU-oriented lattice Boltzmann solver based on a lightweight moment-represented formulation, in which post-collision populations are reconstructed on the fly from a reduced set of moments rather than stored explicitly. This approach significantly lowers the memory footprint, enabling large three-dimensional simulations within the constraints of modern accelerator architectures, where VRAM capacity and bandwidth are critical resources. The method is assessed through standard single- and two-component benchmarks demonstrating good accuracy and stability. Extensive scaling experiments on multi-GPU systems show near-ideal weak scaling up to 512 GPUs and sustained performance across different velocity sets. The combination of reduced memory usage, competitive throughput, and stable energy efficiency makes the proposed formulation a practical route for large-scale lattice Boltzmann simulations on current and emerging HPC platforms.
\end{abstract}


\begin{keyword}
Computational fluid dynamics \sep Lattice Boltzmann \sep HCP GPU computing \sep Multiphase flows
\end{keyword}

\end{frontmatter}


\section{Introduction}

Over the last three decades, the lattice Boltzmann method (LBM) has become a well-established tool for computational fluid dynamics, thanks to its kinetic foundation, algorithmic simplicity, and remarkable suitability for parallel computing \cite{tiribocchi2025lattice,succi2018lattice,kruger2017lattice,ladd1994numerical}. In its standard form, LBM stores the full set of discrete particle populations at every lattice site. This makes the method inherently memory intensive, especially for three-dimensional simulations on modern GPU-based architectures \cite{tran2017performance}. This issue has long been recognized as one of the main practical limitations of large-scale LB simulations \cite{lehmann2022esoteric,succi2019towards}.

The idea that the full set of populations may not always need to be stored explicitly is not entirely new. Early reflections in this direction can already be found in the work of Ladd and Verberg \cite{ladd2001lattice}, who pointed out that substantial memory savings might be achieved by exploiting the relation between distribution functions and hydrodynamic fields. In a broader sense, this observation is rooted in the very structure of the lattice Boltzmann method: the discrete populations are not arbitrary variables, but a finite kinetic representation whose physically relevant content is encoded in a limited set of moments, namely density, momentum, and the tensors governing the stress and higher-order non-equilibrium contributions \cite{kruger2017lattice}.

A decisive theoretical step came with the Hermite-based reformulation of lattice Boltzmann models and with the development of regularized schemes \cite{latt2006lattice,latt2007hydrodynamic,montessori2015lattice}. In the work of Latt and Chopard \cite{latt2006lattice}, the method was recast so as to emphasize macroscopically relevant variables rather than the microscopic populations themselves, thereby clarifying which degrees of freedom are essential to recover the Navier--Stokes dynamics. This viewpoint was later strengthened by regularization procedures, where the non-equilibrium part of the distribution is projected onto a controlled Hermite subspace, filtering out spurious high-order contributions and improving both stability and accuracy \cite{wissocq2022hydrodynamic,wissocq2020linear,feng2019hybrid,nathen2018stability,jacob2018new,mattila2017high,coreixas2017recursive,malaspinas2015increasing}. In this context, the Hermite representation provides a compact way of organizing the kinetic hierarchy in terms of moments. For the athermal case, this structure becomes especially powerful because the equilibrium Hermite coefficients satisfy a simple recursive relation, allowing higher-order coefficients to be reconstructed from lower-order moments in a compact and physically consistent way \cite{shan2006kinetic,malaspinas2015increasing}.

These developments opened the way to what is now commonly called \emph{moment-represented} or \emph{lightweight} lattice Boltzmann. The central idea is simple: instead of storing all lattice populations, one stores only the hydrodynamically relevant moments---typically density, momentum, and the second-order tensor associated with the non-equilibrium stress---while reconstructing the populations on the fly whenever they are needed for collision and propagation. In three dimensions, this reduces the number of stored variables per lattice site from, for example, 19 populations in a D3Q19 formulation to only 10 moment variables. Since contemporary high-performance implementations of LBM are often limited more by memory bandwidth than by floating-point throughput \cite{lehmann2022esoteric,succi2019towards}, such a reduction can translate into a substantial performance gain.

The modern formulation of moment-represented LBM has been developed primarily in the context of regularized schemes. Vardhan \emph{et al.} \cite{vardhan2019moment} demonstrated that storing only moment-based data can significantly reduce memory footprint and data motion on massively parallel architectures, while preserving the accuracy of regularized LBM for fluid-dynamical applications. This line of work was further extended by Gounley \emph{et al.}  \cite{gounley2021propagation}, who introduced a dedicated propagation pattern for the moment representation, showing that careful control of cache reuse and data movement is essential to fully exploit the potential of the compressed kinetic description. More recently, Ferrari \emph{et al.} \cite{ferrari2023graphic} provided a three-dimensional GPU implementation of the moment representation, confirming that the reduction in global memory traffic can yield substantial speedups with no appreciable loss of accuracy, especially in single precision.

Within this line of research, the work by Tiribocchi \emph{et al.} \cite{tiribocchi2023lightweight} offers a particularly clear synthesis of both the historical motivations and the computational implications of the moment representation strategy, also called \emph{lightweight} lattice Boltzmann. The formulation in Ref. \cite{tiribocchi2023lightweight} revisits earlier theoretical ideas in light of modern hardware constraints and shows that, in regularized multicomponent lattice Boltzmann models, reconstructing populations from hydrodynamic information may lead to large savings in memory occupation and data-access costs without compromising the underlying physics. In this sense, lightweight LBM is not merely a technical optimization, but the natural outcome of a mature understanding of which kinetic degrees of freedom are genuinely needed by the numerical scheme and which can instead be reconstructed when required.

In this work, we present \textit{LBFAST}, a high-performance lattice Boltzmann solver specifically developed to implement the \emph{lightweight} lattice Boltzmann approach on modern GPU-accelerated supercomputers. Within this framework, \textit{LBFAST} follows the \emph{moment-represented} lattice Boltzmann paradigm, in which the post-collision populations are reconstructed on the fly from a reduced set of hydrodynamic and kinetic variables and immediately streamed to neighboring lattice nodes to update the reduced set of variables at the new time step, without ever being stored as a full population array in global device memory \cite{tiribocchi2023lightweight,vardhan2019moment,gounley2021propagation,ferrari2023graphic}.
\textcolor{black}{Compared with the previous lightweight implementation of Tiribocchi et al. \cite{tiribocchi2023lightweight}, the present work introduces two main advances. First, \textit{LBFAST} extends the population-free, moment-represented strategy to a distributed-memory multi-GPU solver supporting one-, two-, and three-dimensional MPI domain decompositions. Second, the underlying multiphase formulation is changed from the weakly compressible color-gradient setting, in which two component distribution functions are reconstructed separately, to an incompressible velocity--pressure formulation coupled to a conservative Allen--Cahn equation for the phase field \cite{lauricella2025acclb}. In this setting, density and viscosity are reconstructed from the phase field, while pressure and velocity are evolved through the lightweight moment-represented flow solver.}
\textcolor{black}{Further, the present work extends this strategy to a distributed-memory multi-GPU setting and assesses it on a modern GPU-accelerated supercomputer through numerical validation and strong- and weak-scaling measurements using up to 512 GPUs.}

This strategy is particularly well suited to modern accelerator-based architectures, where lattice Boltzmann performance is typically limited by both memory bandwidth and memory footprint. In this context, the moment-represented, \emph{lightweight} LBM formulation implemented in \textit{LBFAST} provides a versatile framework for next-generation high-performance fluid dynamics solvers, especially in applications involving multicomponent physics, large three-dimensional domains, and GPU acceleration, where the available device memory is a key constraint.

\section{Methods}
\label{sec:methods}

\textit{LBFAST} is formulated within the velocity--pressure framework already introduced in Refs. \cite{montessori2026breakdown,lauricella2025thread,dinesh2019phase,fakhari2017improved,zu2013phase} , in which the flow solver evolves the velocity field \(u_\alpha\) and the pressure  \(p\), while the density is not obtained from the zeroth moment of the flow populations. Instead, both density and viscosity are reconstructed from the phase field \(\phi\), which is transported by a conservative Allen--Cahn equation. In this setting, the governing equations read
\begin{subequations}\label{eq:NS}
\begin{align}
   \partial_t \rho(\phi) + \partial_\alpha \left(\rho(\phi) u_\alpha\right) &= 0,
   \label{eq:NSa} \\
   \partial_t \left(\rho(\phi) u_\alpha\right)
   + \partial_\beta \left(\rho(\phi)\, u_\alpha u_\beta\right)
   &=
   -\partial_\alpha p
   + \partial_\beta
   \left[
      \rho(\phi)\,\nu(\phi)
      \left(
         \partial_\beta u_\alpha + \partial_\alpha u_\beta
      \right)
   \right]
   + F_\alpha .
   \label{eq:NSb}
\end{align}
\end{subequations}
where $\nu$ denotes the kinematic viscosity, $\rho$ the density and \(F_\alpha\) is the total force including the surface tension. Greek indices denote Cartesian tensor components, and summation over repeated indices is implied.

The local density and kinematic viscosity are interpolated from the phase field through linear blending laws, $\rho(\phi)=\rho_L+\phi \left(\rho_H - \rho_L\right)$, and $\nu(\phi)=\nu_L+\phi \left(\nu_H - \nu_L\right)$, 
where $\phi=1$ and $\phi=0$ correspond to the heavy and light phases, respectively.

The interface dynamics is described by a conservative Allen--Cahn equation,
\begin{equation} \label{ACeq}
   \partial_t  \phi
    +
    u_\alpha \partial_\alpha \phi
    =
    D \, \partial_\alpha \partial_\alpha \phi
    -
    \lambda \, \partial_\alpha \bigl( \phi (1 - \phi) n_\alpha \bigr),
\end{equation}
where $D$ is the interface diffusivity and $\lambda = 4D/W$, with $W$ denoting the interface thickness and the term $-\lambda \, \partial_\alpha \bigl( \phi (1 - \phi) n_\alpha \bigr)$ acting as an interface compression contribution, counterbalancing numerical diffusion and preserving a sharp interface profile \cite{lauricella2025thread}. The unit normal to the interface is defined as $n_\alpha = \partial_\alpha \phi/|\nabla \phi|$.

In the lightweight moment-represented Lattice Boltzmann (LLB) approach, Eqs \ref{eq:NSa} and \ref{eq:NSb} are indirectly solved integrating the Lattice Boltzmann equation \cite{feng2019hybrid,feng2019hybrid2}:
\begin{equation} \label{pullLB}
f_i(x_\alpha,t + \Delta t) = f^{eq}_{i}(x_\alpha-c_{i\alpha}\Delta t,t) + (1-\omega)f^{neq}_{i}(x_\alpha-c_{i\alpha}\Delta t,t) + \frac{1}{2}S_i(x_\alpha-c_{i\alpha}\Delta t,t) ,
\end{equation}
where collision and streaming are combined into a single update step within a pulled scheme, $S_i$ denotes the forcing term and $f^{neq}_{i}=f_i-f^{eq}_{i} + \frac{1}{2} S_i$
ensures the second-order accuracy in time. The relaxation time $\tau=1/\omega$ determines the kinematic viscosity through $\nu = c_s^2(\tau - 0.5)$ with $c_s$ the lattice speed of sound \cite{succi2018lattice,kruger2017lattice}.
However, rather than evolving the full set of populations, in LLB the distribution functions $f_i$ are reconstructed on the fly as a truncated Hermite expansion \cite{shan2006kinetic}:
\begin{subequations}\label{eq:reqtfeq}
\begin{align}
f_i^{eq}&=w_i \sum_n \frac{1}{c_s^{2n} n!} \mathcal{H}^{(n)}_{i\alpha_1...\alpha_n}a^{(n)}_{eq,\alpha_1...\alpha_n},\label{eq:reqfeq}\\
f_i^{neq}&=w_i \sum_n \frac{1}{c_s^{2n} n!} \mathcal{H}^{(n)}_{i\alpha_1...\alpha_n}a^{(n)}_{neq,\alpha_1...\alpha_n},
   \label{eq:reqfneq}
\end{align}
\end{subequations}
with the Hermite coefficients obtained by projection:
\begin{subequations} \label{hermite_teq}
\begin{align}
a^{(n)}_{eq,\alpha_1,...,\alpha_n}&=\sum_i \mathcal{H}^{(n)}_{i\alpha_1,...,\alpha_n} f_i^{eq},\label{hermite_eq}\\
a^{(n)}_{neq,\alpha_1,...,\alpha_n}&=\sum_i \mathcal{H}^{(n)}_{i\alpha_1,...,\alpha_n} f_i^{neq}.\label{hermite_neq}
\end{align}
\end{subequations}

In practice, the expansion is truncated at second order, $n=2$, obtaining the total Hermite coefficient as $a^{(n)}_{\alpha_1,...,\alpha_n}=a^{(n)}_{eq,\alpha_1,...,\alpha_n}+a^{(n)}_{neq,\alpha_1,...,\alpha_n}$, consistently with the recovery of the Navier--Stokes equations. The discrete Hermite basis reads
$H^{(0)}_{i} = 1$, 
$H^{(1)}_{i\alpha} = c_{i\alpha}$, 
$H^{(2)}_{i\alpha\beta} = (c_{i\alpha} c_{i\beta} - c_s^2\delta_{\alpha\beta})$.
Under this truncation, and within a velocity-based formulation \cite{lauricella2025thread,dinesh2019phase,fakhari2017improved,zu2013phase}, the equilibrium coefficients are explicitly expressed in terms of the primitive variables:
\begin{subequations}
\begin{align}
a^{(0)}_{eq}=p^*, \quad a^{(1)}_{eq,\alpha}=u_\alpha, \quad a^{(2)}_{eq,\alpha\beta}=u_\alpha u_\beta, \\
a^{(0)}_{neq}=0, \quad a^{(1)}_{neq,\alpha}=0, \quad a^{(2)}_{neq,\alpha\beta}=-\frac{c^s_2}{\omega}\left(
         \partial_\beta u_\alpha + \partial_\alpha u_\beta
      \right)+
\frac{1}{2\rho}
\left(
F_\alpha u_\beta + F_\beta u_\alpha
\right),
\end{align}
\end{subequations}
where $p^*$ denotes a rescaled pressure variable, defined as $p^* = p/(\rho(\phi) c_s^2)$, allowing the equilibrium to be formulated directly in terms of the primitive variables, i.e., the velocity and the pressure fields (rather than the momentum), and the non-equilibrium contribution is entirely contained in the second-order tensor $a^{(2)}_{neq,\alpha\beta}$, related to the viscous stress \cite{malaspinas2015increasing,kruger2009shear} and the force term arising from the shifted definition $f^{neq}_{i}=f_i-f^{eq}_{i} + \frac{1}{2} S_i$. This projection effectively filters out non-hydrodynamic degrees of freedom, a key ingredient of regularized formulations \cite{latt2006lattice}.

In the LLB approach \cite{tiribocchi2023lightweight},  exploiting Eqs \ref{eq:reqtfeq} and \ref{hermite_teq}, the lattice Boltzmann Eq. \ref{pullLB} is recast as:
\begin{equation}
\label{pullLLB}
\begin{aligned}
a^{*(n)}_{\alpha_1,\ldots,\alpha_n}(x_\alpha,t + \Delta t)
&=
\sum_i \mathcal{H}^{(n)}_{i\alpha_1,\ldots,\alpha_n}
\Biggl[
w_i \sum_m \frac{1}{c_s^{2m} m!}
\mathcal{H}^{(m)}_{i\alpha_1,\ldots,\alpha_m}
\\
&\qquad \times
\Bigl(
a^{(m)}_{eq,\alpha_1,\ldots,\alpha_m}(x_\alpha-c_{i\alpha}\Delta t,t)
+
(1-\omega)\,
a^{(m)}_{neq,\alpha_1,\ldots,\alpha_m}(x_\alpha-c_{i\alpha}\Delta t,t)
\Bigr)
\\
&\qquad +
\frac{1}{2}S_i(x_\alpha-c_{i\alpha}\Delta t,t)
\Biggr].
\end{aligned}
\end{equation}
where $i$ runs over all discrete lattice directions of the adopted velocity set, while $m$ spans the Hermite expansion order from $0$ to $2$.
In order to remain consistent with the definition $f^{neq}_{i}=f_i-f^{eq}_{i} + \frac{1}{2} S_i$, the Hermite coefficients are updated as:
\begin{equation}\label{eq:corr}
a^{(0)}=a^{*(0)}, \quad a^{(1)}_{\alpha} =
a^{*(1)}_{\alpha}
+\frac{F_\alpha}{2\rho}, \quad a^{(2)}_{\alpha\beta}=a^{*(2)}_{\alpha\beta}+\frac{1}{2\rho}
\left(
F_\alpha u_\beta + F_\beta u_\alpha
\right),
\end{equation}
with the force term computed at the updated time, $t + \Delta t$.

Thus, the LLB approach consists of integrating in time Eq.~\ref{pullLLB} and applying the correction in Eq. \ref{eq:corr}, which depends only on the Hermite coefficients at the previous time step. The discrete forcing term $S_i$ is defined according to the Guo scheme \cite{guo2002discrete}:
\begin{equation} \label{guoforce}
S_i = w_i \left( \frac{c_{i\alpha} - u_\alpha}{c_s^2} + \frac{c_{i\beta} u_\beta}{c_s^4} c_{i\alpha} \right)\frac{F_\alpha}{\rho},
\end{equation}
where $F_\alpha$ denotes the total force per unit volume.
In particular, the vector $F_\alpha$ can be split into four contributions, namely $F_\alpha=F_\alpha^s+ F_\alpha^b + F_\alpha^p  +F_\alpha^\nu$. 
Following Refs. \cite{lauricella2025thread,dinesh2019phase,jacqmin2000contact}, the capillary force is expressed as $F_\alpha^s = \mu_\phi \, \partial_{\alpha} \phi$,
where $\mu_\phi
=
4\beta\,\phi(\phi-1)\left(\phi-\frac12\right)
-\kappa\nabla^2\phi$ is the chemical potential,
with the coefficients chosen so as to recover the prescribed surface tension $\sigma$ and interface thickness $W$, yielding $\beta = 12 \sigma/W$ and
$\kappa = 3\sigma W/2$. The symbol $F_\alpha^b$ stands for an external body force, whereas $F_\alpha^p$ denotes the pressure force \cite{lauricella2025thread} equal to $F_\alpha^p=-p^*c_s^2 \partial_\alpha \rho$. This choice allows reconstructing $\partial_\alpha  p =\rho c_s^2 \partial_\alpha  p^* + p^* c_s^2 \partial_\alpha  \rho$ in Eq. \ref{eq:NSb} with the first term already embedded in the LB equation \cite{lauricella2025thread}. Lastly, the contribution arising from viscosity gradients across the interface $F_{\alpha}^{\nu}=\nu\,\omega\,c_s^2\,a^{*(2)}_{\mathrm{neq},\alpha\beta}\,\partial_\beta \rho$ accounts for  the non-equilibrium stress-based forcing term, as in Refs.~\cite{lauricella2025acclb,lauricella2025thread}.

It is worth noting that Eq.~\ref{pullLLB} is formulated in a way that is independent of the specific choice of the discrete velocity set, and can therefore be readily implemented with different quadrature rules. In the present work, we consider the D3Q19 and D3Q27 models with the Hermite expansion truncated at second order, together with a higher-order D3Q27h formulation based on the complete Hermite basis. In the latter case, the Hermite coefficients are reconstructed through the recursive closure relations of Shan et al.~\cite{shan2006kinetic}, consistently with the high-order Hermite framework discussed by Malaspinas~\cite{malaspinas2015increasing}.

Spatial derivatives entering both the forcing terms and the phase-field equation are computed using lattice-based stencils constructed from the weights of the D3Q27 velocity set. In particular, the first- and second-order derivatives are approximated as
$\partial_\alpha \Psi = \frac{1}{c_s^2} \sum_i w_i \Psi(x_\alpha + c_{i\alpha})\, c_{i\alpha}$, and
$\partial_\alpha \partial_\beta \Psi = \frac{1}{c_s^2} \left( \sum_{i \neq 0} w_i \Psi(x_\alpha + c_{i\alpha}) - w_0 \Psi(x_\alpha) \right)$.

Given the small time step imposed by the LBM, the conservative Allen--Cahn Eq. \ref{ACeq} is discretized using a forward-time, centered-space (FTCS) scheme. This finite-difference approach ensures stability while reducing computational cost and memory usage.
\section{Implementation}

The code is implemented in \texttt{CUDA Fortran}, with additional \texttt{OpenACC} directives, and uses MPI for distributed-memory parallel execution across multiple GPUs. The domain can be decomposed along one, two, or three directions, allowing flexibility in the mapping of large three-dimensional problems while keeping the collision step strictly local and the streaming step based on the on-the-fly reconstruction.

Inter-process communication is carried out using non-blocking MPI primitives. At each time step, halo layers are first assembled to gather the Hermite coefficients required by neighboring subdomains. Asynchronous send and receive calls are issued, and the computation proceeds on the interior of the subdomains while data transfers are in progress. A final synchronization through \texttt{MPI\_Waitall} ensures completion of communications before updating the boundary nodes. This overlap between communication and computation reduces idle time and improves scalability.

The Hermite coefficients are stored using a block-structured layout. Lattice nodes are grouped into compact three-dimensional tiles mapped onto CUDA thread blocks, and the data are arranged as
\begin{equation}
n_{\mathrm{tot}} = 
TILE\_DIM_x \, TILE\_DIM_y \, TILE\_DIM_z \, n_{\mathrm{hfields}} \, n_{\mathrm{blocks}}
\end{equation}
Two arrays, \texttt{hfields\_flip} and \texttt{hfields\_flop}, are used in a double-buffering scheme. The number of fields is fixed to $n_{\mathrm{hfields}} = 10$, corresponding to the Hermite expansion truncated at second order: one scalar moment $a^{(0)}$, three velocity components $a^{(1)}_{\alpha}$, and six independent components of the symmetric second-order tensor $a^{(2)}_{\alpha\beta}$. This layout is consistent with previous GPU-oriented implementations based on moment representations \cite{ferrari2023graphic}.

Within each tile, threads access data by fixing the local indices $(i,j,k)$, leading to coalesced memory accesses. The non-locality of the streaming step is handled on the fly by reconstructing populations and storing them temporarily in shared memory. The shared buffer includes halo layers and is defined at compile time (e.g., $8\times8\times8$), providing low-latency access and limiting global-memory traffic.

The main kernel follows a fused collide-and-stream approach in a pull formulation. For each lattice site, the Hermite coefficients are read, the populations are reconstructed locally, and their contributions are accumulated directly into the output moments following Eq. \ref{pullLLB}. These output moments (Hermite coefficients) act as local accumulators, so that no full set of populations is stored in global memory. This reduces memory footprint and increases arithmetic intensity.

In this formulation, the macroscopic dynamics is encoded in a reduced set of low-order moments, while higher-order contributions are either filtered, as in regularized approaches \cite{latt2006lattice}, or consistently reconstructed from the retained moments through Hermite-based closures \cite{malaspinas2015increasing,shan2006kinetic}. This separation enables a substantial reduction in memory footprint and allows for a reorganization of the algorithm that is well suited to GPU architectures. The resulting implementation combines a compact data representation with an execution pattern tailored for large-scale simulations. To make the memory saving more explicit, let us consider a cubic domain of size $512^3$, for which the total number of lattice sites is $N = 512^3$. In the present lightweight formulation, the solver stores only the ten second-order Hermite fields and uses a double-buffering strategy, so that the total storage amounts to $2 \times 10 = 20$ scalar fields per lattice site. By contrast, a standard population-based implementation with flip--flop storage requires $2 \times 19 = 38$ scalar fields for a D3Q19 stencil and $2 \times 27 = 54$ scalar fields for a D3Q27 stencil. Hence, the total device memory required by the main lattice arrays is $M = N \, n_{\mathrm{fields}} \, b$,
where $n_{\mathrm{fields}}$ is the number of stored scalar fields per site and $b$ is the size in bytes of a scalar.

In double precision ($b=8$ bytes), this gives exactly $20$~GB for the present moment-represented implementation, compared with $38$~GB for a standard D3Q19 population scheme and $54$~GB for a standard D3Q27 one in a double-buffering approach. Therefore, the lightweight approach reduces the memory occupation by about $47.4\%$ with respect to D3Q19 and by about $63.0\%$ with respect to D3Q27. In single precision, the corresponding values become $10$~GB, $19$~GB, and $27$~GB, respectively, with the same percentage reductions. These figures refer only to the main lattice-resident flow arrays and clearly illustrate the advantage of the moment-represented formulation for large three-dimensional simulations on GPU architectures.
\section{Validation and Benchmarking}
\label{sec:bench}

We first consider a monocomponent benchmark in order to test the accuracy of the solver in a simple and controlled setting. To this end, we study the viscous decay of the three-dimensional Taylor--Green vortex in a fully periodic cubic box of size $512^3$, at fixed density $\rho=1.0$ and kinematic viscosity $\nu=0.04074$. With the characteristic length defined as $L=N/(2\pi)$ and the initial velocity equal to $U_0=0.04$, this choice corresponds to a Reynolds number $Re \simeq 80$.

The initial condition is taken as in Ref.~\cite{wang2013high}, consistently adapted to the pressure-based formulation~\cite{lauricella2025acclb,dinesh2019phase}, where the hydrodynamic pressure is defined up to an arbitrary constant and taken with zero mean. The velocity field is prescribed as
\begin{subequations}\label{eq:tg}
\begin{align}
u_x(x,y,z)&=U_0 \sin(x)\cos(y)\cos(z),\\
u_y(x,y,z)&=-U_0 \cos(x)\sin(y)\cos(z),\\
u_z(x,y,z)&=0,\\
p(x,y,z)&=\frac{U_0^2}{16}\left[\cos(2x)+\cos(2y)\right]\left[2+\cos(2z)\right],
\end{align}
\end{subequations}
where the dimensionless coordinates are defined as
$x = 2\pi (i-1)/N, \quad
y = 2\pi (j-1)/N, \quad
z = 2\pi (k-1)/N$,
and $U_0$ denotes the initial velocity amplitude.

As a diagnostic quantity, we monitor the volume-averaged kinetic energy
$E(t)=\langle u_\alpha u_\alpha \rangle/2$,
and compare its evolution with the analytical viscous decay of the fundamental mode,
$E(t)/E(0)=\exp\left[-6\nu k^2 t\right]$, being $k=2\pi/N$.

The same test is repeated with three different discrete velocity sets: D3Q19, D3Q27 with Hermite expansion truncated at second order, and a high-order D3Q27h model based on the full Hermite basis.

\begin{figure}[ht] \centering \includegraphics[width=0.7\linewidth]{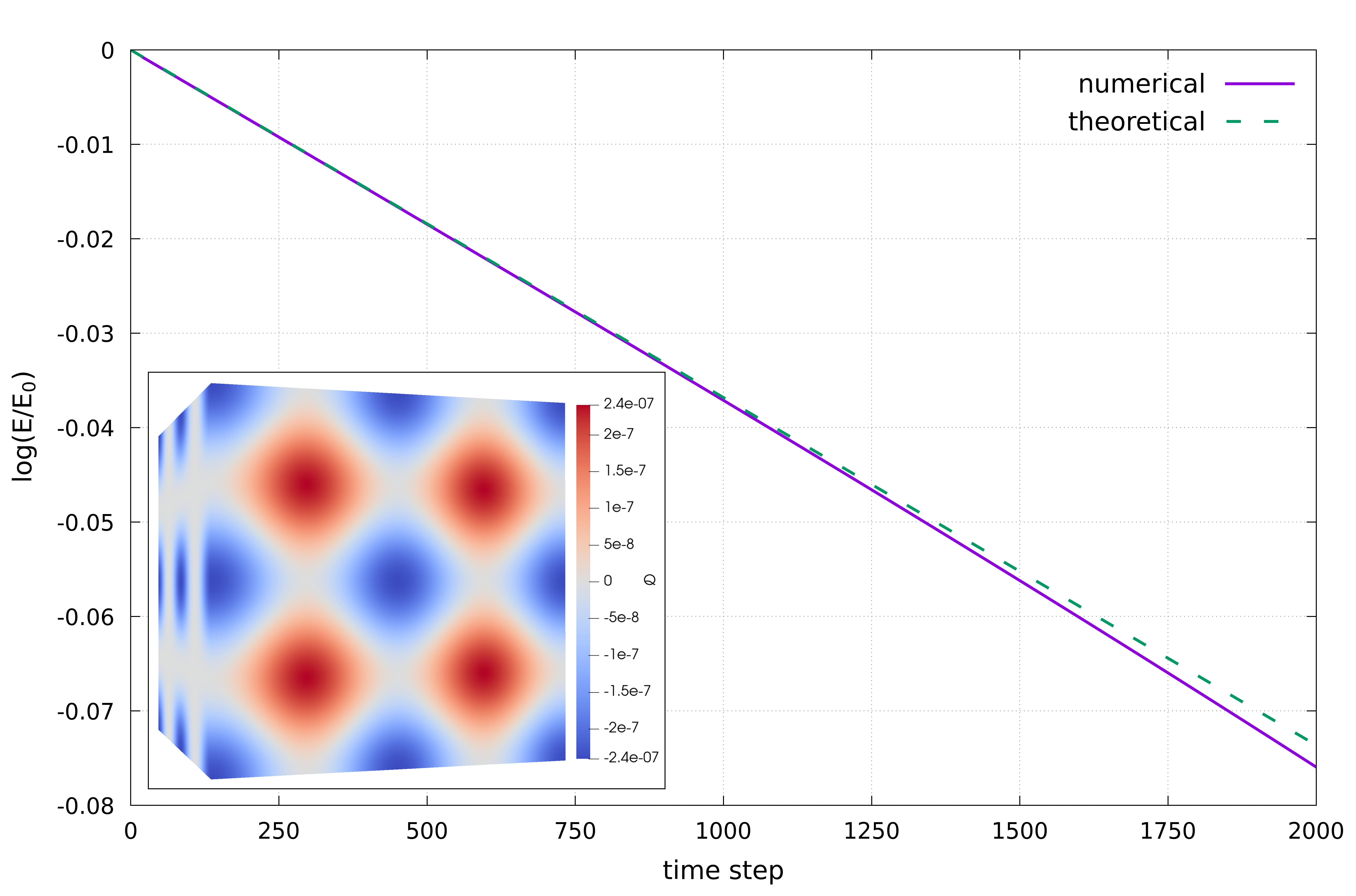} \caption{Decay of the normalized kinetic energy in the Taylor--Green vortex for the D3Q27 scheme. Numerical results are in close agreement with the analytical prediction, yielding $\nu_{\mathrm{fit}}=0.04079$ with a relative error of $\sim 1.3\times 10^{-3}$. Inset: \textcolor{black}{the initial Q-criterion scalar field is mapped onto the surface of the computational domain, with $Q = \frac{1}{2}(\|\boldsymbol{\Omega}\|^2-\|\mathbf{S}\|^2)$, where $\boldsymbol{\Omega}$ and $\mathbf{S}$ are the antisymmetric and symmetric parts of the velocity-gradient tensor, respectively.}}\label{fig:tg} \end{figure}

Figure \ref{fig:tg} reports the temporal evolution of the normalized kinetic energy, $E(t)/E(0)$, for the D3Q27 scheme. The numerical data follow closely the expected exponential decay, with only minor deviations at longer times, where the numerical curve exhibits a slightly faster decay than the theoretical prediction. \textcolor{black}{The slight late-time departure from the theoretical exponential decay is attributed to the accumulation of numerical dissipation and finite-resolution errors, a behavior commonly monitored in vortex-decay benchmarks as a sensitive indicator of the effective viscosity and numerical accuracy of the scheme \cite{malaspinas2009}.}

A linear fit of $\log(E/E_0)$ over the initial decay range yields an effective kinematic viscosity $\nu_{\mathrm{fit}} = 0.04079$, in very good agreement with the input value, with a relative deviation of approximately $1.3\times 10^{-3}$. This confirms that both the kinetic model and its implementation accurately reproduce the viscous decay of the fundamental mode.
An inset shows the initial $Q$-criterion field at $t=0$, highlighting the coherent vortical structures of the Taylor--Green configuration, where $Q = \tfrac{1}{2}\left(\Omega_{\alpha\beta}\Omega_{\alpha\beta}
- S_{\alpha\beta}S_{\alpha\beta}\right)$ denotes the second invariant of the velocity gradient tensor, with $\Omega_{\alpha\beta}$ and $S_{\alpha\beta}$ the antisymmetric and symmetric parts, respectively.

Finally, for completeness, we report that analogous fits yield $\nu_{\mathrm{fit}}=0.040794$ for D3Q19 and $\nu_{\mathrm{fit}}=0.040791$ for the high-order D3Q27h model, with relative deviations of $\sim 1.3\times 10^{-3}$ and $\sim 1.2\times 10^{-3}$, respectively.

We consider the layered Poiseuille flow as a two-phase benchmark with viscosity contrast, to assess the accuracy of the solver in the presence of a diffuse but stationary interface. This configuration provides a standard validation test for binary-fluid lattice Boltzmann models, since an analytical steady solution exists for two immiscible layers under laminar conditions \cite{zu2013phase}. The flow is driven by a constant body acceleration $a_z$ along $z$, with no-slip walls at the boundaries normal to $x$ and periodicity in $y$ and $z$. The domain size is $L_x \times L_y \times L_z = 512 \times 8 \times 1024$, and $a_z = 1.0\times 10^{-7}$.

The two fluids are initially arranged in layers across the channel: the less viscous phase occupies $1 \le x \le L_x/2$, while the more viscous one fills the remaining half. Their kinematic viscosities are $\nu_1 = 0.05$ and $\nu_2 = 0.4$, with densities $\rho_1 = 0.5$ and $\rho_2 = 1.5$. Density and viscosity are reconstructed from the order parameter $\phi$, and the interface is initialized as a diffuse layer of thickness $W=4$ with surface tension $\sigma=0.03$. Capillary effects do not affect the steady velocity profile due to the planar interface.

\begin{figure}[ht] \centering \includegraphics[width=0.7\linewidth]{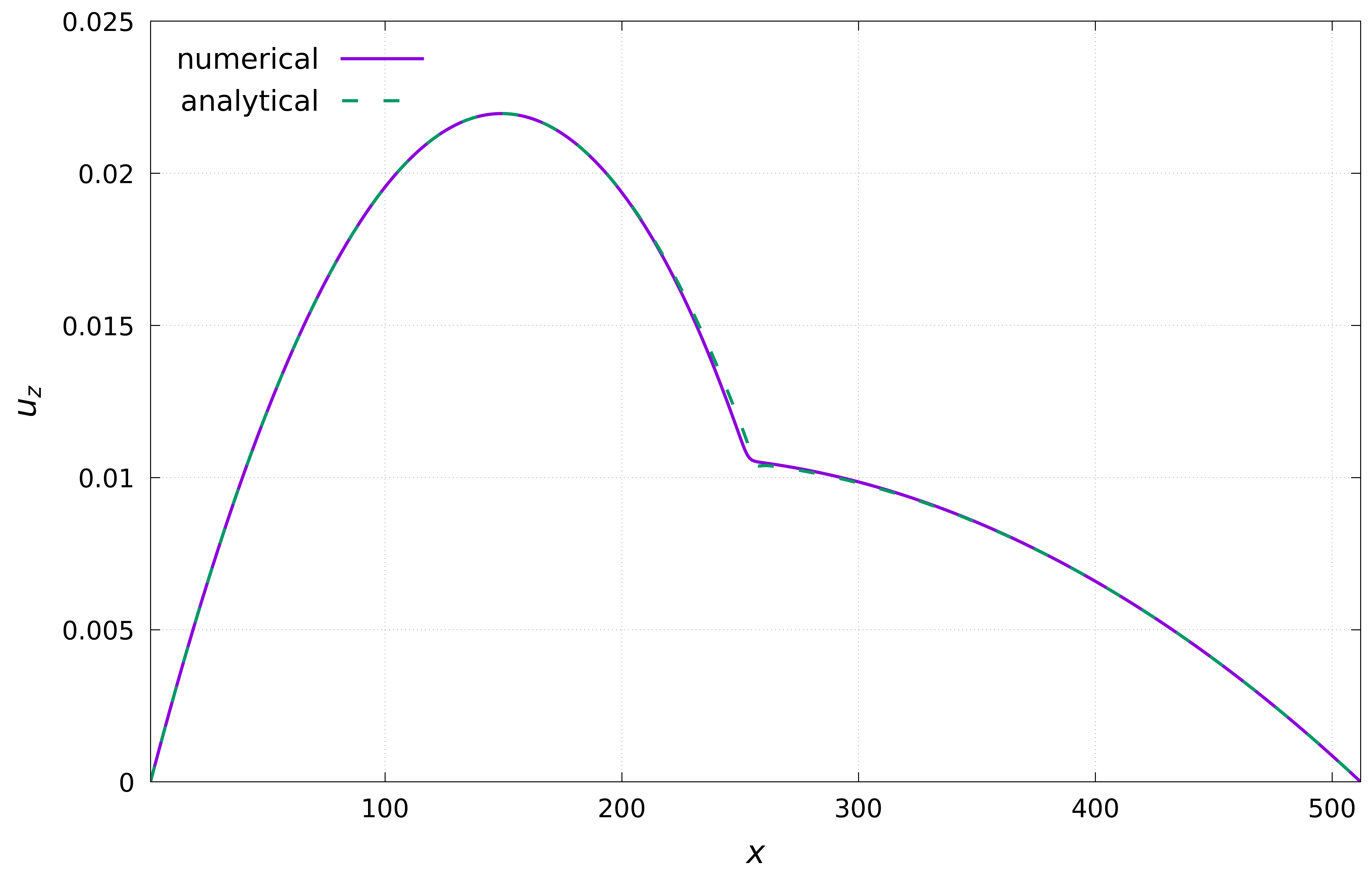}
\caption{Comparison between the numerical and analytical velocity profiles for the layered Poiseuille flow with viscosity contrast for the D3Q27 scheme. Excellent agreement is observed across the entire channel, including the interfacial region where the slope changes due to the viscosity jump.}\label{fig:pois} \end{figure}

Figure~\ref{fig:pois} compares the numerical velocity profile obtained from the D3Q27 scheme with the analytical solution reported in \cite{zu2013phase}. The agreement is very good across the entire channel. A small deviation is observed in the interfacial region, where the slope changes due to the viscosity jump: this discrepancy is expected, as the analytical solution assumes a sharp interface, while the present model employs a diffuse-interface formulation. In particular, the largest relative error is observed at the interface, 
\(x = 251\), and amounts to \(4.18\times10^{-2}\), \(4.18\times10^{-2}\), and \(3.97\times10^{-2}\) 
for the D3Q19, standard D3Q27, and high-order D3Q27h schemes, respectively. Overall, this benchmark confirms that the implementation accurately reproduces the layered Poiseuille solution and correctly captures viscosity contrasts within the diffuse-interface framework.

In addition to the layered Poiseuille configuration, we performed a Laplace test to assess the accuracy of surface tension and isotropy for the different velocity sets. A spherical droplet of radius \(R = 64\) lattice units is initialized at the center of a cubic domain of size \(512^3\), ensuring negligible finite-size effects. The two phases are characterized by equal densities and viscosities, \(\rho_1 = \rho_2 = 1\) and \(\nu_1 = \nu_2 = 0.1\), while the interface is described by a diffuse profile of thickness \(W = 4\) and surface tension \(\sigma = 3.000 \times 10^{-2}\).

At equilibrium, the pressure jump across the interface is expected to follow the Laplace law, \(\Delta P = 2\sigma/R\). The numerical results obtained with the three stencils are in very good agreement with the theoretical prediction. In particular, the measured surface tension is \(\sigma = 2.796 \times 10^{-2} \pm 9.976 \times 10^{-4}\), \(\sigma = 2.796 \times 10^{-2} \pm 9.979 \times 10^{-4}\), and \(\sigma = 2.796 \times 10^{-2} \pm 9.979 \times 10^{-4}\) for the D3Q19, standard D3Q27, and high-order D3Q27h schemes, respectively. The corresponding pressure jumps and radii are also consistent across all cases, with only negligible differences within statistical uncertainty.
These results indicate that all three discretizations provide a consistent estimate of the macroscopic surface tension within statistical uncertainty for this static configuration.
\section{Performance Analysis}
\label{sec:perform}

\begin{table*}[t]
\caption{Strong-scaling with fixed total domain cubic box of side 512 lattice points: GLUPS values for one-component (Taylor--Green benchmark) and two-component (Laplace benchmark) systems in single and double precision.}
\label{tab:strong_all_components}
\centering
\resizebox{0.9\textwidth}{!}{%
\begin{tabular}{r c r r r r r r r r r r r r}
\hline
GPU & decomp & \multicolumn{6}{c}{one-component} & \multicolumn{6}{c}{two-component} \\
\cline{3-8}\cline{9-14}
 &  & \multicolumn{3}{c}{single precision} & \multicolumn{3}{c}{double precision} & \multicolumn{3}{c}{single precision} & \multicolumn{3}{c}{double precision} \\
\cline{3-5}\cline{6-8}\cline{9-11}\cline{12-14}
 &  & {\scriptsize D3Q19} & {\scriptsize D3Q27} & {\scriptsize D3Q27h} & {\scriptsize D3Q19} & {\scriptsize D3Q27} & {\scriptsize D3Q27h} & {\scriptsize D3Q19} & {\scriptsize D3Q27} & {\scriptsize D3Q27h} & {\scriptsize D3Q19} & {\scriptsize D3Q27} & {\scriptsize D3Q27h} \\
\hline
1   & 1x1x1   & 2.3  & 1.8  & 1.0  & 1.4  & 1.0  & 0.4  & 1.6  & 1.4  & 0.8  & 0.8  & 0.7  & 0.2 \\
2   & 1x1x2   & 4.4  & 3.5  & 1.9  & 2.6  & 2.0  & 0.8  & 3.0  & 2.6  & 1.6  & 1.5  & 1.2  & 0.5 \\
4   & 1x1x4   & 8.2  & 6.6  & 3.6  & 5.0  & 3.9  & 1.6  & 5.7  & 4.8  & 3.0  & 2.9  & 2.5  & 0.9 \\
4   & 1x2x2   & 8.2  & 6.6  & 3.6  & 4.9  & 3.9  & 1.6  & 5.6  & 4.8  & 3.0  & 2.9  & 2.4  & 0.9 \\
8   & 1x2x4   & 14.1 & 11.6 & 6.8  & 8.7  & 7.0  & 3.1  & 9.3  & 8.2  & 5.4  & 5.1  & 4.3  & 1.8 \\
8   & 2x2x2   & 13.8 & 11.4 & 6.7  & 8.6  & 6.9  & 3.0  & 9.1  & 8.0  & 5.3  & 5.0  & 4.3  & 1.7 \\
16  & 1x4x4   & 22.3 & 19.1 & 12.0 & 14.7 & 12.1 & 5.8  & 14.1 & 12.6 & 9.1  & 8.4  & 7.4  & 3.3 \\
16  & 2x2x4   & 22.2 & 18.8 & 11.9 & 14.5 & 12.0 & 5.7  & 13.5 & 12.4 & 8.9  & 8.4  & 7.2  & 3.2 \\
32  & 1x4x8   & 31.5 & 28.2 & 19.6 & 21.4 & 18.5 & 10.0 & 18.7 & 16.9 & 13.3 & 11.8 & 10.8 & 5.6 \\
32  & 2x4x4   & 32.4 & 28.5 & 19.8 & 22.3 & 19.2 & 10.2 & 18.6 & 17.2 & 13.4 & 12.3 & 11.1 & 5.7 \\
64  & 1x8x8   & 41.7 & 39.0 & 29.0 & 29.9 & 26.4 & 16.8 & 21.9 & 21.6 & 14.9 & 15.7 & 14.5 & 9.1 \\
64  & 4x4x4   & 39.0 & 40.1 & 27.7 & 30.2 & 26.3 & 17.2 & 19.7 & 20.5 & 17.5 & 14.4 & 14.5 & 9.2 \\
128 & 1x8x16  & 49.5 & 47.6 & 41.5 & 38.5 & 35.4 & 26.4 & 24.4 & 23.6 & 23.4 & 19.3 & 16.5 & 13.2 \\
128 & 4x4x8   & 41.4 & 42.2 & 40.4 & 37.9 & 33.8 & 26.5 & 19.8 & 22.0 & 21.6 & 17.8 & 14.8 & 13.1 \\
256 & 1x16x16 & 44.8 & 58.5 & 47.6 & 46.1 & 43.9 & 34.3 & 21.6 & 26.1 & 24.7 & 19.5 & 21.3 & 17.0 \\
256 & 4x8x8   & 44.3 & 48.5 & 45.1 & 45.1 & 40.7 & 33.8 & 19.1 & 23.9 & 19.0 & 18.1 & 18.0 & 16.8 \\
512 & 1x16x32 & 62.3 & 55.9 & 55.1 & 56.1 & 53.9 & 46.9 & 27.3 & 24.8 & 24.5 & 23.4 & 23.2 & 20.7 \\
512 & 8x8x8   & 48.1 & 46.0 & 44.5 & 48.3 & 47.4 & 42.7 & 21.8 & 19.4 & 21.5 & 20.4 & 19.0 & 20.5 \\
\hline
\end{tabular}%
}
\end{table*}

\begin{table*}[t]
\caption{Weak-scaling with fixed sub-domain cubic box of side 512 lattice points: GLUPS values for one-component (Taylor--Green benchmark) and two-component (Laplace benchmark) systems in single and double precision.}
\label{tab:weak_all_components}
\centering
\resizebox{0.9\textwidth}{!}{%
\begin{tabular}{r c r r r r r r r r r r r r}
\hline
GPU & decomp & \multicolumn{6}{c}{one-component} & \multicolumn{6}{c}{two-component} \\
\cline{3-8}\cline{9-14}
 &  & \multicolumn{3}{c}{single precision} & \multicolumn{3}{c}{double precision} & \multicolumn{3}{c}{single precision} & \multicolumn{3}{c}{double precision} \\
\cline{3-5}\cline{6-8}\cline{9-11}\cline{12-14}
 &  & {\scriptsize D3Q19} & {\scriptsize D3Q27} & {\scriptsize D3Q27h} & {\scriptsize D3Q19} & {\scriptsize D3Q27} & {\scriptsize D3Q27h} & {\scriptsize D3Q19} & {\scriptsize D3Q27} & {\scriptsize D3Q27h} & {\scriptsize D3Q19} & {\scriptsize D3Q27} & {\scriptsize D3Q27h} \\
\hline
1   & 1x1x1   & 2.3   & 1.8   & 1.0   & 1.3   & 1.0   & 0.4   & 1.6   & 1.4   & 0.8   & 0.8   & 0.7   & 0.2 \\
2   & 1x1x2   & 4.5   & 3.6   & 1.9   & 2.6   & 2.1   & 0.8   & 3.2   & 2.7   & 1.6   & 1.6   & 1.3   & 0.5 \\
4   & 1x1x4   & 9.0   & 7.1   & 3.8   & 5.3   & 4.1   & 1.7   & 6.4   & 5.3   & 3.2   & 3.1   & 2.6   & 0.9 \\
4   & 1x2x2   & 9.0   & 7.0   & 3.8   & 5.2   & 4.1   & 1.6   & 6.3   & 5.3   & 3.2   & 3.1   & 2.6   & 0.9 \\
8   & 1x1x8   & 17.7  & 14.0  & 7.5   & 10.3  & 8.1   & 3.3   & 12.5  & 10.5  & 6.3   & 6.2   & 5.1   & 1.9 \\
8   & 1x2x4   & 17.4  & 13.8  & 7.4   & 10.2  & 8.0   & 3.3   & 12.2  & 10.2  & 6.3   & 6.0   & 5.0   & 1.8 \\
8   & 2x2x2   & 17.4  & 13.8  & 7.4   & 10.1  & 8.0   & 3.3   & 12.1  & 10.2  & 6.3   & 6.0   & 5.0   & 1.8 \\
16  & 1x1x16  & 35.3  & 27.8  & 14.9  & 20.7  & 16.2  & 6.6   & 24.9  & 20.8  & 12.7  & 12.3  & 10.2  & 3.7 \\
16  & 1x4x4   & 34.7  & 27.6  & 14.8  & 20.2  & 15.9  & 6.5   & 24.3  & 20.3  & 12.5  & 12.0  & 10.0  & 3.6 \\
16  & 2x2x4   & 33.6  & 26.7  & 14.6  & 19.4  & 15.4  & 6.4   & 23.3  & 19.8  & 12.3  & 11.5  & 9.6   & 3.6 \\
32  & 1x1x32  & 70.9  & 55.9  & 29.9  & 41.2  & 32.4  & 13.1  & 49.6  & 41.6  & 25.2  & 24.6  & 20.3  & 7.4 \\
32  & 1x4x8   & 67.9  & 54.2  & 29.5  & 39.3  & 31.2  & 12.7  & 47.3  & 39.9  & 24.7  & 23.4  & 19.5  & 7.1 \\
32  & 2x4x4   & 66.7  & 53.3  & 29.2  & 38.4  & 30.6  & 12.8  & 46.1  & 39.1  & 24.3  & 22.9  & 19.2  & 7.1 \\
64  & 1x1x64  & 140.7 & 111.7 & 59.9  & 82.3  & 64.5  & 26.0  & 98.3  & 83.6  & 50.6  & 48.9  & 40.8  & 14.8 \\
64  & 1x8x8   & 134.8 & 107.7 & 58.7  & 78.4  & 62.3  & 25.7  & 93.8  & 79.6  & 49.3  & 46.7  & 39.1  & 14.3 \\
64  & 4x4x4   & 132.2 & 106.2 & 58.3  & 76.7  & 60.8  & 25.3  & 91.2  & 78.2  & 48.6  & 45.7  & 37.9  & 14.1 \\
128 & 1x1x128 & 278.8 & 223.2 & 119.6 & 164.8 & 128.9 & 52.1  & 196.7 & 165.9 & 100.8 & 97.8  & 81.0  & 29.7 \\
128 & 1x8x16  & 268.0 & 215.9 & 117.7 & 157.2 & 124.4 & 51.5  & 187.4 & 157.9 & 98.3  & 93.0  & 77.2  & 28.5 \\
128 & 4x4x8   & 256.8 & 208.7 & 115.2 & 150.2 & 119.9 & 50.7  & 179.2 & 152.4 & 95.9  & 88.8  & 74.8  & 28.3 \\
256 & 1x1x256 & 564.3 & 447.5 & 239.5 & 329.8 & 258.7 & 104.4 & 393.8 & 332.2 & 201.9 & 195.1 & 162.0 & 59.3 \\
256 & 1x16x16 & 536.0 & 428.7 & 234.7 & 311.1 & 247.5 & 102.4 & 370.9 & 317.6 & 196.2 & 184.9 & 154.4 & 56.9 \\
256 & 4x8x8   & 512.6 & 413.1 & 230.3 & 296.1 & 237.5 & 101.1 & 351.7 & 303.1 & 191.2 & 176.7 & 148.7 & 57.5 \\
512 & 1x1x512 & 1131.4 & 890.2 & 479.2 & 658.9 & 517.3 & 209.4 & 767.8 & 641.1 & 386.3 & 379.3 & 312.6 & 115.6 \\
512 & 1x16x32 & 1001.5 & 797.7 & 430.5 & 582.2 & 457.3 & 191.5 & 709.3 & 596.2 & 362.7 & 351.5 & 290.8 & 111.2 \\
512 & 8x8x8   & 960.5 & 770.7 & 421.1 & 556.2 & 440.6 & 187.1 & 674.2 & 566.5 & 353.4 & 334.3 & 277.2 & 108.2 \\
\hline
\end{tabular}%
}
\end{table*}

In this section, we assess the performance and energy efficiency of the present implementation on large-scale GPU systems. 
All simulations have been carried out on the \textit{Leonardo} supercomputer at CINECA \cite{turisini2024leonardo}. \textcolor{black}{Each Leonardo
compute node is equipped with four NVIDIA A100 GPUs, each with 64 GB of HBM2 memory; hence,
the largest calculations reported in this work, using 512 GPUs, correspond to 128 compute nodes.} 
The code is compiled with the NVIDIA HPC SDK and executed using CUDA-aware MPI.

Performance is quantified in terms of GLUPS (Giga Lattice Updates Per Second), defined as the total number of lattice updates performed per second, expressed in billions. 
\begin{equation}
\text{GLUPS}=\frac{L_x L_y L_z}{10^9 t_{\text{s}}},
\end{equation}
where $t_{\text{s}}$ is the run (wall-clock) time (in seconds) per single time step iteration, while $L_x$, $L_y$, and $L_z$ are the domain sizes.
This metric provides a direct measure of the solver throughput and is used throughout this section to compare different configurations. To evaluate scalability, we also introduce the parallel efficiency, defined as
\begin{equation}
\mathrm{PE}(N) = \frac{\mathrm{GLUPS}(N)}{N \, \mathrm{GLUPS}(1)},
\end{equation}
where $\mathrm{GLUPS}(N)$ is the performance obtained using $N$ GPUs.

\textcolor{black}{A direct performance comparison with the standard population-based formulation was already reported in our previous lightweight LBM study \cite{tiribocchi2023lightweight}. There, a two-component D3Q19 color-gradient benchmark in single precision on an NVIDIA Tesla V100 was used to compare the population-free lightweight implementation with the corresponding standard CGLB formulation. The lightweight version \cite{tiribocchi2023lightweight} reached about $0.622$ GLUPS, compared with $0.277$ GLUPS for the population-based implementation, yielding a speed-up of about $2.25$. The same test showed a reduction of the GPU memory footprint from $1.1411$ GB to $0.6782$ GB, corresponding to about $40\%$ memory saving.}

\textcolor{black}{The present work builds on those single-GPU results and addresses the complementary problem of deploying the lightweight moment-represented strategy in a distributed-memory multi-GPU setting. In \textit{LBFAST}, the reduced moment fields are stored using a flip--flop double-buffering layout, while full population arrays are never stored in global device memory. The analysis below therefore focuses on numerical validation, MPI communication, one-, two-, and three-dimensional domain decomposition, and strong and weak scaling up to $512$ GPUs.}

Strong- and weak-scaling results are reported in Tables~\ref{tab:strong_all_components} and \ref{tab:weak_all_components} for both the single-component (Taylor--Green) and two-component (Laplace) benchmarks as introduced in Section \ref{sec:bench}, considering different velocity sets and numerical precisions.

In strong scaling, the solver exhibits good parallel efficiency up to several hundred GPUs, with a gradual saturation at the largest scales. This behavior is mainly due to the increasing impact of communication costs as the local sub-domain size decreases. The choice of the velocity set significantly affects performance: D3Q19 achieves the highest GLUPS, followed by D3Q27, while D3Q27h is consistently more expensive due to the larger number of degrees of freedom and reconstruction operations. The same trend is observed in both single- and two-component systems and in both precisions.

\begin{figure}[ht] \centering \includegraphics[width=0.7\linewidth]{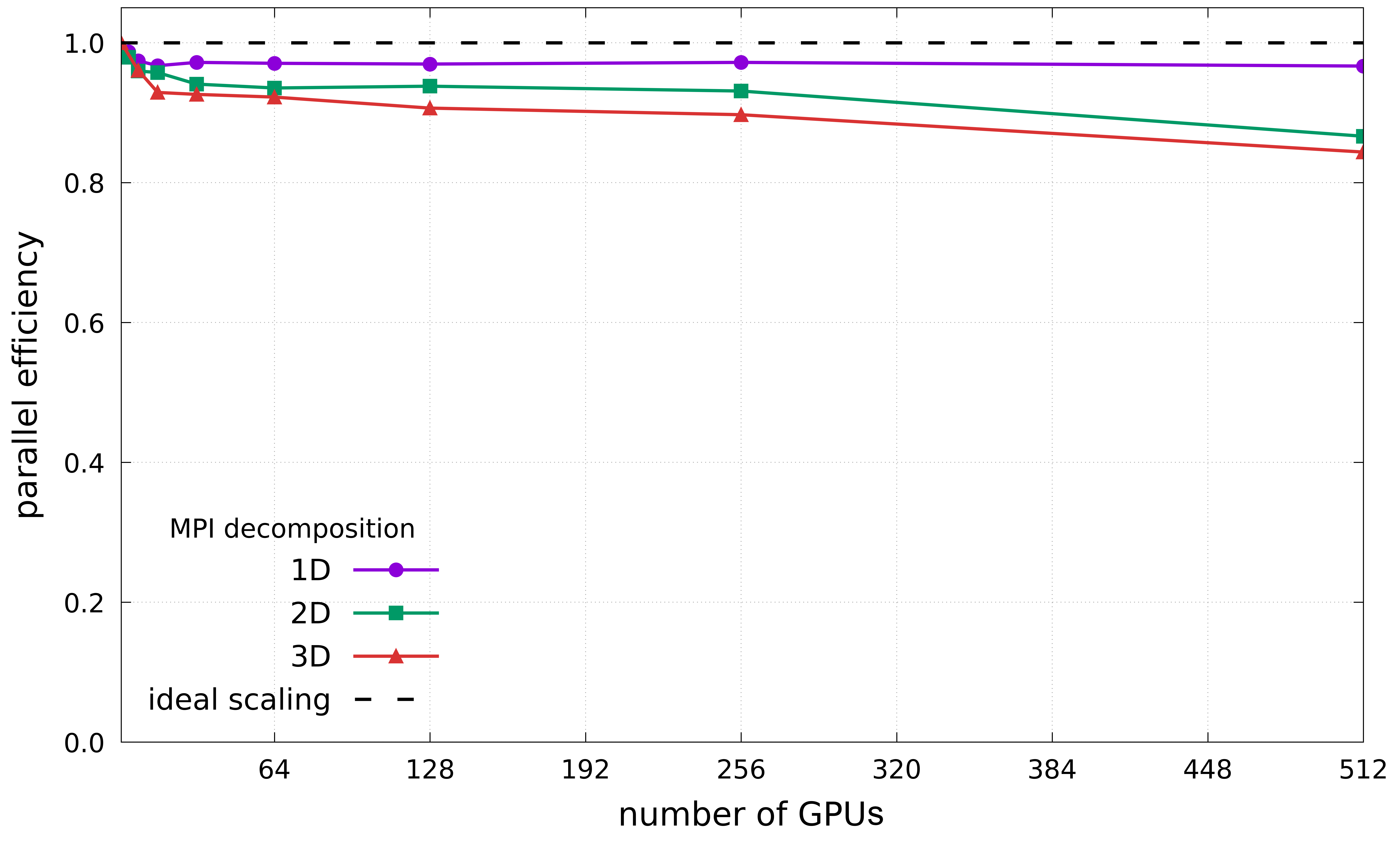}
\caption{Weak scaling parallel efficiency for the single-component Taylor--Green benchmark (Section~\ref{sec:bench}) using the D3Q27 velocity set. Results are shown for different MPI decompositions (1D, 2D, 3D), reporting for each GPU count the best-performing MPI configuration for the 2D and 3D decompositions. The dashed line indicates ideal scaling.}\label{fig:pe} \end{figure}

\begin{figure}[ht] \centering \includegraphics[width=0.7\linewidth]{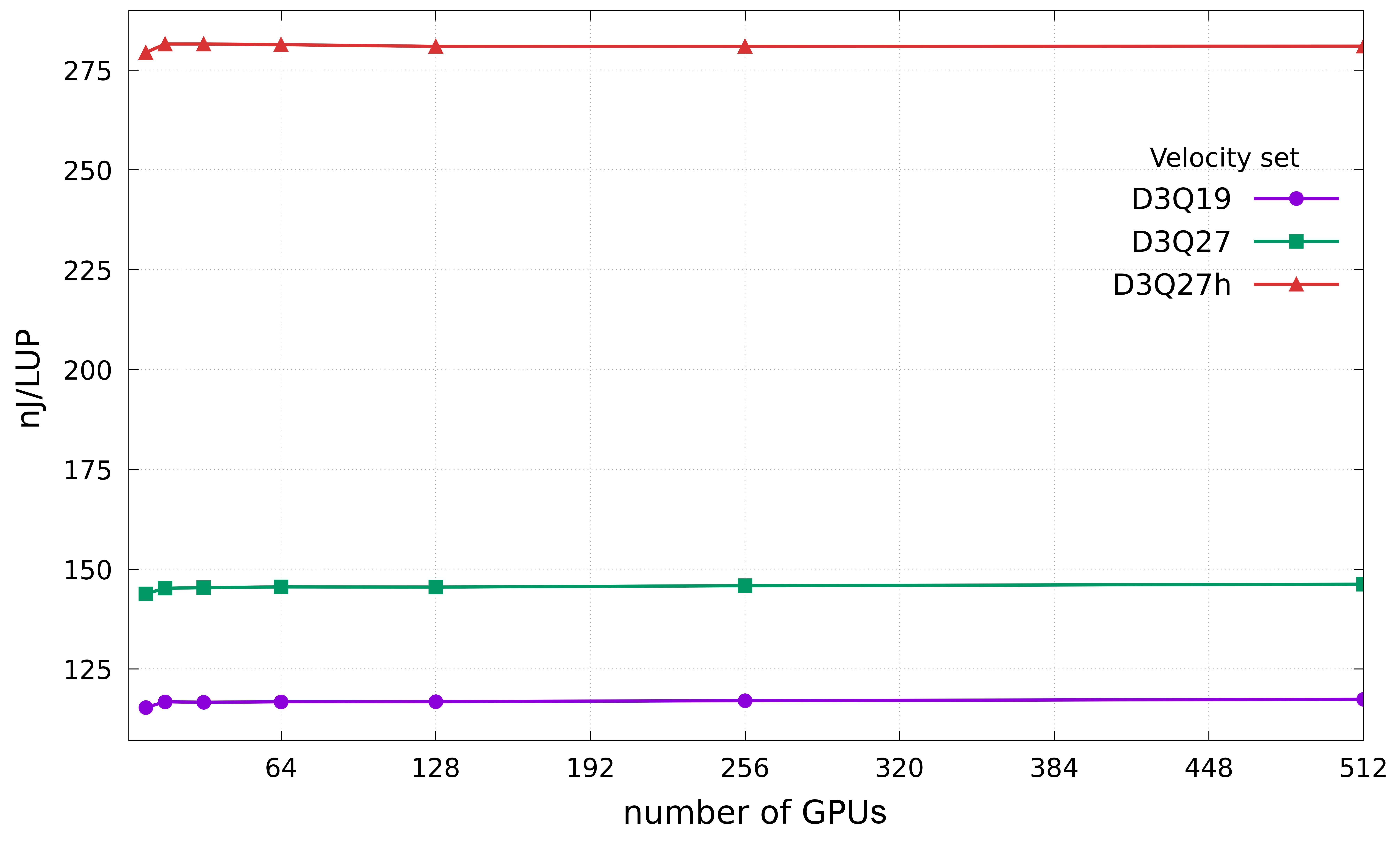}
\caption{Energy per lattice update (nJ/LUP) in weak scaling for the single-component Taylor--Green benchmark (Section~\ref{sec:bench}), using a 3D MPI decomposition. Comparison among different velocity sets (D3Q19, D3Q27, and D3Q27h).}\label{fig:energy} \end{figure}

In weak scaling, the code remains close to ideal up to 512 GPUs for all configurations, with only a modest loss of efficiency at the largest scales, suggesting that communication is largely hidden by computation. Figure~\ref{fig:pe} shows the parallel efficiency for the D3Q27 case across different MPI decompositions.
The 1D layout performs slightly better over the whole range, while 2D and 3D decompositions show a more visible, though still limited, degradation as the number of GPUs increases. This trend reflects the growing communication burden associated with higher-dimensional layouts: in the present setup, each subdomain exchanges data with two neighbors in 1D, eight in 2D, and up to twenty-six in 3D, resulting in progressively higher communication costs.

The comparison between one- and two-component systems reveals a systematic reduction in GLUPS for the latter. This is expected, as the phase-field evolution and the computation of interfacial forces introduce additional memory accesses and arithmetic operations. The performance gap becomes more pronounced in double precision, where memory bandwidth limitations are more severe.

Energy consumption has been measured using the NVIDIA Management Library (NVML), by computing the difference of the cumulative energy counter provided by \texttt{nvmlDeviceGetTotalEnergyConsumption} between the beginning and the end of each run. Based on this, we define the energy per lattice update as
\begin{equation}
\mathrm{nJ/LUP} = \frac{E_{\mathrm{tot}}}{N_{\mathrm{LUP}}},
\end{equation}
where $E_{\mathrm{tot}}$ is the total energy consumed (in nanojoule) and $N_{\mathrm{LUP}} = L_x L_y L_z \, N_t$,
with $N_t$ the number of time steps.

The behavior of this metric in weak scaling is reported in Figure~\ref{fig:energy} for several velocity sets. The results show that the energy per lattice update remains approximately constant as the number of GPUs increases, indicating that the solver preserves its energy efficiency at scale. As for performance, the velocity set plays a key role: D3Q19 is the most energy-efficient configuration, while D3Q27h exhibits the highest energy cost per update, reflecting the increased computational workload associated with the larger number of degrees of freedom and reconstruction operations.

Overall, the results indicate that the present implementation achieves good scalability on modern GPU architectures, while maintaining a stable balance between performance and energy consumption across a wide range of problem sizes.

\textcolor{black}{We finally position the present performance results with respect to state-of-the-art standard LB solvers for multiphase flows on GPU infrastructures. }

\textcolor{black}{In the class of distribution-function-based multiphase LBM solvers,
accLB~\cite{lauricella2025acclb} provides a relevant multi-GPU reference.
It couples a conservative Allen--Cahn phase-field formulation with a
regularized LBM and relies on hybrid MPI--OpenACC parallelism. The accLB code
demonstrates strong and weak scaling up to 64 GPUs, reaching sustained
throughputs above 150 GLUPS for a $512^3$ lattice domain; on a single
NVIDIA A100 GPU, it achieves 2.58 GLUPS in single precision. 
Another state-of-the-art multiphase LBM reference is the conservative Allen--Cahn implementation of Holzer et al. \cite{holzer2021highly}, integrated into the waLBerla framework through automatic code generation. Their solver targets immiscible fluids at high density ratios, employs optimized generated CPU/GPU kernels, and exhibits excellent weak scaling on Piz Daint up to $2048$ GPUs, with a reported parallel efficiency of about $98\%$.
In particular, Holzer et al.~\cite{holzer2021highly} reported $2.66$ GLUPS for the
phase-field kernel and $1.36$ GLUPS for the hydrodynamic kernel on a
single NVIDIA Tesla V100 in double precision.}

\textcolor{black}{However, a quantitative one-to-one comparison among these solvers is not straightforward, since they differ not only in hardware and programming model, but also in the underlying LB formulation: accLB \cite{lauricella2025acclb} uses a regularized population-based approach, whereas the waLBerla implementation of Holzer et al. \cite{holzer2021highly} relies on distribution-function-based Allen--Cahn LB steps with BGK/MRT-type collisional dynamics. These differences affect memory traffic, arithmetic intensity, number of stored fields, stencil choice, and communication volume, and therefore make raw GLUPS values only partially comparable.}

\textcolor{black}{More broadly, the present work also belongs to the current effort of moving CFD solvers to GPU-accelerated architectures. Relevant examples include high-performance DNS/CFD codes such as AFiD~\cite{Zhu_AFiD_2018}, STREAmS~\cite{Bernardini2021}, CaNS~\cite{costa2021}, FluTAS~\cite{crialesi2023flutas}, CaNS-Fizzy~\cite{lupo2025}, the pseudo-spectral DNS solver ported by A. Roccon in Ref.~\cite{Roccon2024}, and URANOS-2.0~\cite{devanna2025}, as well as GPU-oriented LBM codes such as TLBfind~\cite{pelusi2022tlbfind,pelusi2023analysis} and LBcuda~\cite{Bonaccorso2022}. These codes define the broader GPU-CFD landscape, but they are not direct algorithmic counterparts of \textit{LBFAST}, since they generally rely on different discretizations, physical models, or implementation strategies.}

\section{Conclusions}

In this work, we have presented \textit{LBFAST}, a lattice Boltzmann solver designed to efficiently implement the lightweight, moment-represented formulation on modern GPU-based architectures. By avoiding the storage of full population arrays and reconstructing the distributions on the fly from a reduced set of variables, the method significantly reduces the memory footprint while preserving the accuracy of the hydrodynamic description. This feature is particularly relevant for large three-dimensional simulations, where memory bandwidth and device memory capacity are the primary limiting factors.

The numerical benchmarks considered in this study, including the viscous decay of the Taylor--Green vortex and the Laplace test for multicomponent systems, confirm the correctness and robustness of the implementation. In both cases, the solver reproduces the expected physical behavior across a wide range of parameters, demonstrating that the lightweight formulation can be effectively extended to complex flows without compromising stability.

From the performance standpoint, the code exhibits good strong- and weak-scaling properties on the Leonardo supercomputer, maintaining high throughput up to several hundreds of GPUs. The results highlight the importance of the MPI domain decomposition, with multi-dimensional layouts providing a clear advantage at large scales. The comparison among different velocity sets shows a consistent trade-off between computational cost and physical fidelity, with higher-order models requiring additional operations but remaining competitive in terms of scalability.

Energy efficiency has been assessed through the energy per lattice update, showing that the solver preserves a stable energy footprint in weak scaling. This indicates that the increase in parallelism does not introduce additional overheads in terms of energy consumption per unit of work, confirming the effectiveness of the communication--computation overlap and the overall balance of the implementation.

Overall, the present results demonstrate that the lightweight lattice Boltzmann approach provides a viable and efficient framework for large-scale simulations of fluid flows on GPU-accelerated systems. The combination of reduced memory usage, good scalability, and stable energy efficiency makes \textit{LBFAST} a promising tool for next-generation high-performance applications, including multicomponent and thermally fluctuating systems.

\section*{Data availability}
\textcolor{black}{The \textit{LBFAST} source code is publicly available at
\url{https://github.com/lauricella/LBFAST.git}. The software is released under
the Non-Commercial Research License -- Version 1.0, allowing use, copy, and
modification for research, educational, and other non-commercial purposes only.}

\section*{Acknowledgements}

A.M. and M.L. acknowledge funding from the Italian Government through the PRIN grant MOBIOS (ID: 2022N4ZNH3, CUP: F53C24001000006). M.L. and A.T. acknowledge support from GNFM-INdAM. M.L. and S.S. acknowledge support from the European Research Council (ERC Proof of Concept Grant No.~101187935, LBFAST). We acknowledge CINECA for the computing resources provided through the ISCRA-B project MIPLAST (HP10BZY7BK) on the Leonardo Booster system, which is part of the EuroHPC Joint Undertaking (EuroHPC JU) infrastructure.


\end{document}